# A STUDY ON POYNTING EFFECT IN BRAIN WHITE MATTER: A HYPERELASTIC 3D MICROMECHANICAL MODEL

**Mohit Agarwal and Assimina A. Pelegri[1]**

Mechanical and Aerospace Engineering
Rutgers, The State University of New Jersey
Piscataway, NJ, USA

**ABSTRACT**

Shear and torsional load on soft solids such as brain white matter exhibit the so-called Poynting Effect. It is a typical nonlinear phenomenon associated with soft materials whereby they tend to elongate (positive Poynting effect) or contract (negative Poynting effect) in a direction perpendicular to the shearing or twisting plane. In this paper, a novel 3D micromechanical Finite Element Model (FEM) has been developed to depict the Poynting effect in bi-phasic Representative volume element (RVE) with axons embedded in surrounding extra-cellular matrix (ECM) for simulating the brain white matter response under simple and pure shear. In the proposed 3D FEM, nonlinear Ogden hyper-elastic material model describes axons and ECM materials. The modeled bi-phasic RVEs have axons tied to surrounding matrix. In this proof-of-concept (POC) FEM, three simple shear loading configurations and a pure shear scenario were simulated. Root mean square deviation (RMSD) were computed for stress and deformation response plots to depict role of axon-ECM orientations & loading condition on the Poynting effect. Variations in normal stresses (S11, S22, or S33) perpendicular to the shear plane emphasized role of fiber-matrix interactions. At high strains, the stress-strain% plots also indicated modest strain stiffening effects and bending stresses in purely sheared axons.

**Keywords**: Micromechanics, Poynting effect, Numerical modelling, FEM, TBI, Axonal injury, Mechanics of CNS white matter, multi-scale simulation, hyper-elastic materials, matrix-fiber interactions, Python, Abaqus.

**NOMENCLATURE**

| | |
|---|---|
| α | alpha |
| μ | shear moduli (hyper-elastic: Ogden model) |
| λ | principal stretches |
| σ | principal stress |
| τ | shear stress |
| W | Ogden strain-energy density function |
| F | Deformation gradient tensor |
| C | Cauchy-Green deformation tensor (right) |

## 1.   INTRODUCTION

Mechanical response characterization and computational and mathematical modeling of brain tissues have gathered tremendous interest from researchers to understand their role in physiological and pathological conditions such as brain development, brain injury, and surgical procedures [1].

Saccomandi et al. [1] point out that most proposed constitutive models lack general consensus despite all the efforts in modeling brain white matter (BWM). From a characterization point of view, experimental tests setup are often designed to use simple homogenous deformation in specimens such as simple tension or compression, simple shear, torsion, or bi-directional homogenous deformations. The difficulty with such approaches lies in the impossibility of using dog bone-shaped test specimens to realize homogenous deformations over a substantial strain range due to brain tissues' fragility and tackiness. To overcome these limitations, *Destrade et al*. [2] research regarding different characterization simple shear and torsion tests at quasi-static rates emerged as suitable candidates to obtain linear shear-stress and shear-strain relationships over a significant range of shear (up to 60%). These results showed much promise and agreed with previous studies [3, 4].

---

[1] Corresponding author pelegri@rutgers.edu

Similarly, torque-angular displacement linear relationships were discovered by Shuck and Advani [5] and later confirmed by *Balbi* [6] and *Destrade* [7]. While compression tests can achieve homogenous deformation only up to 10% strain, after which the specimen bulges out. In contrast, simple shear tests have worked well up to 45° tilt angles attaining more than 60% stretch. Similar success has been reported with torsion tests of brain white matter.

The natural question arises why we are interested in simple shear and torsion. The reason is to study the underlying *Poynting effect*, a typical nonlinear phenomenon displayed by soft tissues such as brain white matter. When such materials are sheared or twisted, they tend to either elongate (positive Poynting effect) or contract (negative Poynting effect) [8] in a direction perpendicular to the shearing (rectangular specimen) or twisting plane (for cylindrical specimen) [9]. However, due to the lack of shearing devices capable of accurately quantifying and measuring these residual normal forces, further research is required to develop methodologies to help determine brain white matter properties. This is where the proposed research steps in to characterize and model brain white under shead loads. Shear deformation effects have been studied as external traumatic load (TBI case) and surgical tool-brain tissue interaction forces to evaluate resultant forces in hyper-elastically modeled brain white matter and its physiological impact on brain tissues. Although, accurate measurement of the normal forces to characterize brain material properties remains a major challenge.

In this paper, a novel 3D micromechanical Finite Element Model (FEM) has been developed in-house to depict the Poynting effect in the bi-phasic Representative volume element (RVE). Previous research was utilized to determine the suitable volume fraction (VF) of axons embedded in the extra-cellular matrix (ECM) in white matter to model the bi-phasic RVEs for simulating the mechanical response under simple shear loads. Studies indicate that Ogden and Mooney-Rivlin models [10, 11] have shown good agreement between experimental and numerical results to estimate normal forces (N) to depict Poynting effect in soft solids. In this research, a nonlinear Ogden hyper-elastic constitutive material model has been deployed to characterize the bi-phasic RVE in 3D FEM. The brain central nervous system (CNS) comprises of white and gray matter. White matter comprises of myelin coated axons. Axons are long slender projections of neurons which relay information to other neurons, muscles and glands [12, 13]. These axons are embedded in extra cellular matrix (ECM). In the proposed 3D FEM, axon-ECM comprise the bi-phasic constituents. For this proof-of-concept (POC) 3D RVE model, axon-ECM have been assumed to have pure affine boundary condition (i.e., axon is tied down to the ECM) to capture axon-ECM interaction effects [14] on resultant Poynting behavior.

In total, three different simple shear cases: 1) shear along axon length, 2) shear perpendicular to axon fiber and 3) shear across axon fibers were simulated to quantify degree of Poynting behavior. A pure-shear RVE model was also analyzed to understand the role of fiber-matrix interactions on brain's mechanical response to shear (refer **Figure 1**). Root mean square deviation (RMSD) were computed between normal stress-strain response plots for the different simple shear configurations in the ensemble to depict role of axon-ECM orientations trends on dominant Poynting effect directions and resultant normal displacement (U). Variations in degree of normal stresses (S11, S22, or S33) perpendicular to the shear plane in the analyzed three simple shear cases emphasized the role of fiber-matrix interactions and VFs (fiber dispersion) on the Poynting effect [15]. At high strains, strain stiffening effects were also evaluated from the stress-strain plots for the Ogden-modeled bi-phasic RVEs.

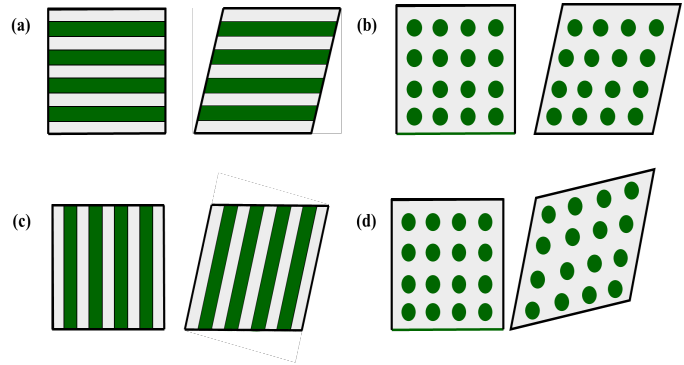

***Figure 1:*** *Schematics of (a) longitudinal shear (along axon fiber axis), (b) shear across fiber (transverse shear), (c) shear perpendicular to axon fibers, and (d) pure shear strain case where shear strain is applied across both opposing axonal fiber faces. The dark green lines represent the axon fibers in (a) and (c), while the dark-green circles depict the axon fiber cross-sections in (b) and (d).*

## 2. MATERIALS AND METHODS

### 2.1 Micromechanical Finite Element Model

The microscale 3D FEM is developed in-house using Abaqus 2020 and Python scripting. Representative Volume Elements (RVEs) are formulated in single RVE units and compounded configurations to simulate simple shear in various configurations. In the current FEM, the average axonal diameter is modeled to be at 0.4395 µm (see **Figure 2**).

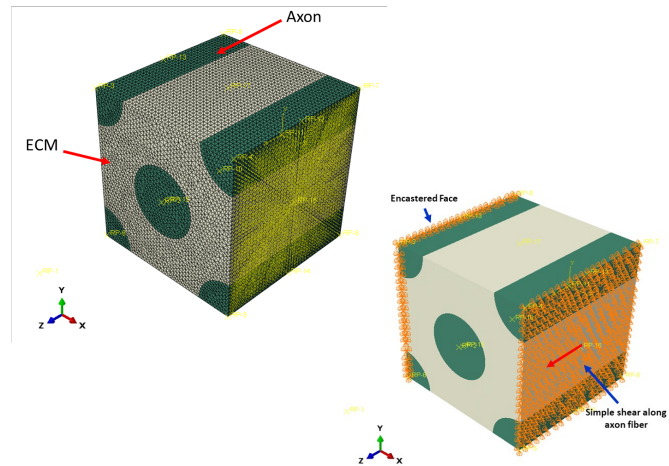

***FIGURE 2:*** *(a) Single RVE FE Model of the ECM and axon assembly packing two axon fibers, volume fraction = 0.3. The ECM is a light beige, and the axon fibers are dark green. (b) Boundary conditions for single RVE. The opposite surface is encastered (fixed) for simple shear, while the sheared face has the strain applied (along the axon fiber shown here). Contact between axons and ECM is defined to be fixed (tied) for this proof-of-concept FEM.*

In the proposed 3D FEM, the axons and the ECM material characterizations are described by non-linear Ogden hyper-elastic constitutive material model. The choice of Ogden hyper-elastic model was motivated by its capability to accurately simulate large quasi-static strains (up to 100%) in very soft-rubber-like materials.

### 2.2 Hyper-elastic (HE) Material Model Component

Nonlinear hyper-elastic models are continually used for simulating soft biological tissues [16-20]. In this research, the Ogden hyper-elastic (HE) material model is used to simulate the ECM and the axons [21] because its nonlinear response allows for more accurate neural tissue characterization at large deformations and strains while capturing the rate-dependent behavior. The Ogden hyper-elastic model is based on three principal stretches, $\lambda_1$, $\lambda_2$, $\lambda_3$, and $2N$ material constants. The strain energy density function, $W$, for the Ogden material model (Equation 1) in Abaqus is formulated as [21, 22]:

$$W = \sum_{1}^{N} \frac{2\mu_i}{\alpha_i^2} (\bar{\lambda}_1^{\alpha_i} + \bar{\lambda}_2^{\alpha_i} + \bar{\lambda}_3^{\alpha_i} - 3) + \sum_{1}^{N} \frac{1}{D_i} (J_{el} - 1)^{2i} \tag{1}$$

where $\bar{\lambda}_i = J^{-\frac{1}{3}} \lambda_i$ and $\overline{\lambda_1 \lambda_2 \lambda_3} = 1$. Here, $J$ signifies a local change of volume and is related to the determinant of the deformation gradient tensor $F$, via the right Cauchy-Green tensor ($C = F^T F$) as $J^2 = det\,(F)^2$, $\mu_i$ represents shear moduli, while $\alpha_i$ (stiffening parameter) and $D_i$ are material parameters. In Equation 1, $J_{el}$ is the elastic volume ratio. The first and the second terms represent the deviatoric and hydrostatic components of the strain energy function. The parameter $D_i = \frac{2}{K_0}$, allows for the inclusion of compressibility where $K_0$ is the initial bulk modulus. The same single parameter Ogden hyper-elastic material is used in this study as well [23]. Therefore, N = 1. Incompressibility implies that $J_{el} = 1$ and is specified in Abaqus by setting $D_1 = 0$. As a result, Abaqus eliminates the hydrostatic component of the strain energy density equation, and the expression reduces to the following:

$$W = \sum_{1}^{N} \frac{2\mu_i}{\alpha_i^2} (\bar{\lambda}_1^{\alpha_i} + \bar{\lambda}_2^{\alpha_i} + \bar{\lambda}_3^{\alpha_i} - 3) \tag{2}$$

For the Ogden HE model, three principal Cauchy stresses values are derived by differentiating $W$ with respect to the extension $\lambda$. If an incompressible material is assumed to be subjected to uniaxial tension ($\sigma_y = \sigma_z = 0$). Then, the corresponding hyper-elastic constitutive model principal stress $\sigma_{uniaxial}$, can then be represented as:

$$\sigma_{uniaxial} = \frac{2\mu}{\alpha} [\lambda^\alpha - \left(\frac{1}{\sqrt{\lambda}}\right)^\alpha] \tag{3}$$

In brain white matter, undulation often prevents axons from experiencing full tensions until a threshold strain limit is attained and then undulation for axon becomes 1. In this paper, as a proof-of-concept FEM, axons are modeled as straight fibers and subjected to various shear modes. The constitutive equations describing resultant normal and shear stresses are explained in §2.3.

For the proposed FEM, The mechanical properties of the axon and glia are assumed from previous work and literature data [12, 13, 17, 18, 24]. The values for shear modulus for the axons and ECM are derived from research by Wu et al. [25], while $\alpha$ is based on the model developed by Meaney [16]. The shear modulus of the ECM is assigned relative to the shear modulus of the axon, considering axons are three times stiffer than ECM, as reported by Arborgast and Marguile's published work [26]. The same methodology has been deployed to model incompressibility for the HE material modeling component [21, 23].

**Table 1** summarizes the material properties $\mu$, $D$ and $a$, and element definition used in the general static simulation of single RVE- HE FE model.

*Table 1: HE Material properties summary of single-RVE FEM*

| Component | $\mu$ | $D$ | $\alpha$ | Element Type |
|---|---|---|---|---|
| | MPa | 1/ MPa | | |
| **Axon** | 2.15E-03 | 0 | 6.19 | C3D8H, C3D4H |
| **ECM** | 8.5 E-04 | 0 | 6.19 | C3D4H |

## 2.3 Analytical modeling Poynting effect

Analytical modeling for describing the Poynting effect has been attempted using various constitutive models such as Neo-Hookean, Mooney-Rivlin, St Venant-Kirchhoff, Blatz-Ko, polynomial, exponential, Ogden, and some even utilized linear elastic material (generalized Hooke's Law) to describe Poynting effect [27]. This study will use the Ogden material model to derive constitutive relationships to analyze Poynting behavior in brain white matter. The formulations described only focus on highlighting key relationships, and the in-depth continuum mechanical model is out of the scope of this paper. Research from *Misra et al.* [27], *Rashid et al.* [10], and a textbook from *Gurtin* [28] can be referenced for a more elaborate discussion.

For a simple shear case (refer to **Figure 3**), let us assume that $\kappa$ is the amount of shear and $\gamma$ is the shear angle. Thus, shear strain $\kappa = \tan(\gamma)$. Let **y** represent the deformed configuration of the brain white matter RVE, which is initially positioned at **X.** Thus, the simple shear equation can be written as:

$$\mathbf{y} = (X_1 + \kappa X_2)\mathbf{e_1} + X_2 \mathbf{e_2} + X_3 \mathbf{e_3} \tag{4}$$

In **Equation 4**, $\{\mathbf{e_1}, \mathbf{e_2}, \mathbf{e_3}\}$ denote the rectangular cartesian coordinate system basis vectors. This equation signifies the shear displacement boundary condition while preventing displacement in the normal direction. The resultant deformation gradient tensor, **F**, can be computed as (**Equation 5**):

$$F = \frac{\partial \mathbf{y}}{\partial \mathbf{X}} = \begin{bmatrix} 1 & \kappa & 0 \\ 0 & 1 & 0 \\ 0 & 0 & 1 \end{bmatrix} \tag{5}$$

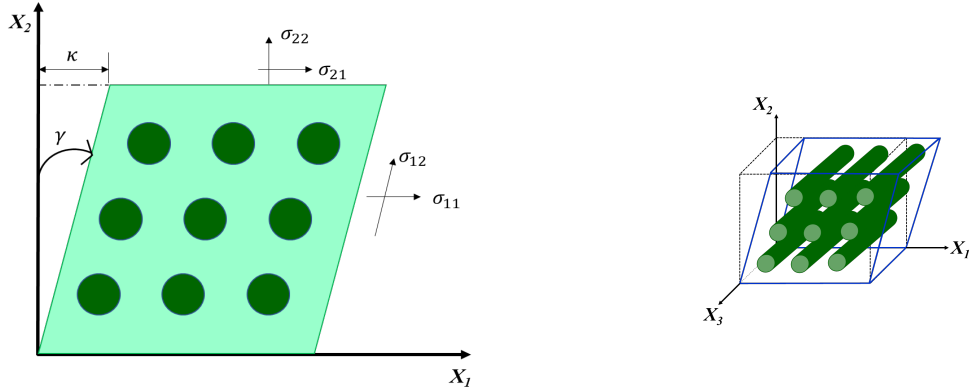

**FIGURE 3:** *(a) Schematic diagram analytically representing simple shear; the shear strain is denoted as $\kappa$ along $X_1$ direction. (b) Simple shear deformation. Dotted lines: sample in the undeformed configuration; solid lines: deformed sample [2].*

For soft tissues undergoing very large strains, nonlinear elastic models (such as hyperelastic (HE) constitutive models) adequately describe the resultant deformation mechanism. For HE materials, Cauchy stress tensor (**Equation 11**) is derived as discussed in §2.2. It is well known that Ogden model were originally designed to model rubber-like materials, but for reasons described in previous section, they can be sufficiently used to highlight nonlinear material behavior for Poynting effect.

$$C = F^T F = \begin{bmatrix} 1 & \kappa & 0 \\ \kappa & 1 + \kappa^2 & 0 \\ 0 & 0 & 1 \end{bmatrix} \tag{6}$$

In general, Ogden hyper-elastically modeled soft tissues' strain energy density function $W$ is a function of two principal inavriants only $W = W(I_1, I_2)$, where $I_1 = tr(C)$, $I_2 = \frac{1}{2}(I_1^2 - tr(C^2))$. But in case of simple shear, $I_1 =$ $I_2 = 3 + \kappa^2$. $C$ is derived as shown in **Equation 6**. The shear component of the Cauchy stress tensor is described as (refer to **equation 7**):

$$\sigma_{12} = 2\kappa \left(\frac{\partial W}{\partial I_1} + \frac{\partial W}{\partial I_2}\right) \tag{7}$$

Thus, $W = W(3 + \kappa^2, 3 + \kappa^2) \equiv \widehat{W}(\kappa)$. Refer to Rashid et al. [10] for more details on the constitutive derivations. Soft tissues are often modeled well by Ogden formulation, and a lot of mechanical test data for brain white matter tissues fits the Ogden hyperplastic (HE) function, hence backing its selection as a constitutive model for this research. The one-term Ogden HE function is given by (refer to **equation 8**):

$$W = \frac{2\mu}{\alpha^2}(\lambda_1^\alpha + \lambda_2^\alpha + \lambda_3^\alpha - 3) \tag{8}$$

In **equation 8**, $\lambda_i$ are the principal stretch ratios (the square roots of the eigenvalues of C). In simple shear models, the principal stretch ratios can be computed as follows:

$$\lambda_1 = \frac{\kappa}{2} + \sqrt{1 + \frac{\kappa^2}{4}}, \lambda_2 = \lambda_1^{-1} = -\frac{\kappa}{2} + \sqrt{1 + \frac{\kappa^2}{4}}, \lambda_3 = 1 \tag{9}$$

Using **Equation 9** in **Equation 8**, we get the resultant expressions for $W$ and $\sigma_{12}$ as depicted in **Equation 10** and **Equation 12**.

$$\widehat{W}(\kappa) = \frac{2\mu}{\alpha^2}\left[\left(\frac{\kappa}{2} + \sqrt{1 + \frac{\kappa^2}{4}}\right)^\alpha + \left(-\frac{\kappa}{2} + \sqrt{1 + \frac{\kappa^2}{4}}\right)^\alpha - 2\right] \tag{10}$$

Using the generalized Cauchy Stress tensor formula for isotropic homogenous hyperelastic materials of the Ogden form (see **equation 11**), the normal and shear stress components can be represented as shown in **Equations 13, 14,** and **12**.

$$\sigma_{ij} = -p\delta_{ij} + \lambda_i \frac{\partial W}{\partial \lambda_j} \tag{11}$$

As seen in **Equation 14**, $\sigma_{22}$ is non-zero. The presence of this non-zero normal stress ($\sigma_{22}$), and the inequality $\sigma_{11} \neq \sigma_{22}$ is a manifestation of the Poynting effect and is a result of material non-linearities in hyper elastically modeled brain white matter soft tissues [27].

$$\sigma_{12} = \widehat{W}'(\kappa) = \frac{\mu}{\alpha} \frac{1}{\sqrt{1 + \left(\frac{\kappa^2}{4}\right)}}\left[\left(\frac{\kappa}{2} + \sqrt{1 + \frac{\kappa^2}{4}}\right)^\alpha - \left(-\frac{\kappa}{2} + \sqrt{1 + \frac{\kappa^2}{4}}\right)^\alpha\right] \tag{12}$$

$$\sigma_{11} = \frac{2\mu}{\alpha}\left[\left(\frac{\kappa}{2} + \sqrt{1 + \frac{\kappa^2}{4}}\right)^\alpha - 1\right] \tag{13}$$

$$\sigma_{22} = T_{22} = \frac{2\mu}{\alpha}\left[\left(\frac{\kappa}{2} + \sqrt{1 + \frac{\kappa^2}{4}}\right)^{-\alpha} - 1\right] \tag{14}$$

In their study on the one-term isotropic Ogden model for soft tissues [29], Horgan and Murphy share insights on nonlinear elastic materials and the gap between current analytical models and actual physical realization by considering three boundary condition scenarios. They point out that extra Ogden invariants could help attain a more rational and reliable prediction of the Poynting effect for complex geometries and boundary conditions. In their analytical model, they derive an expression for traction ($T$) and shear forces ($S$) That arises on the inclined faces to maintain the state of simple shear. Their analysis revealed that a theoretical optimal value of $\alpha = 6$ (in Equation 8) could negate the need to impose any normal traction forces ($N$) on the inclined faces to analytically realize simple shear for one-term Ogden hyperelastic soft tissues. Such a configuration would manifest a moderate tensile hydrostatic stress ($H$) and moderate positive normal stress ($T_{22}$) (same as $\sigma_{22}$ in this paper) for moderate amounts of shear, it also proved the computational feasibility of the shear strain limits explored in the proof-of-concept FEM model described in later sections.

$$T_{33} = \sigma_{33} = \frac{2\mu}{\alpha}\left(\frac{(\lambda^\alpha - 1)(\lambda^{6-\alpha} - 1)}{1 + \lambda^6}\right) \tag{15}$$

$$H = \frac{2\mu}{\alpha}\left(1 + \frac{\lambda^{\alpha+6} - 2\lambda^\alpha + \lambda^{-\alpha} - 2\lambda^{6-\alpha}}{1 + \lambda^6}\right) \tag{16}$$

Following the case of zero normal traction on the inclined faces from Horgan's research [29, 30], out-of-plane stress ($T_{22}$). For the Ogden modeled HE brain soft tissues, resultant out-of-plane stress ($T_{33}$) and the hydrostatic stress ($H$) are obtained as depicted in **equation 15** and **Equation 16** (Note: notations $T_{33}$ signify $\sigma_{33}$ in this paper). For convenience, let us set $\lambda_1 \equiv \lambda$ and $\lambda_2 = 1/\lambda_1$. While exact congruence between analytical models and computational results is difficult, the proposed Ogden model for simulating Poynting effect can perform reasonably well to yield reasonable stress profiles. In future, even higher order invariants in Ogden model and boundary-layer approach could be explored on defining constraints near lateral and inclined faces of the RVEs to overcome ambiguities in formulation of simple shear models for incompressible materials.

**Note:** From a computational framework perspective, in numerical tools such as Abaqus (General, explicit) solver, these constitutive equations should not be misinterpreted as homogenized representations for entire RVEs. Instead, the constitutive equations (4) to (16) are computed for both axon and ECM phase elements (using corresponding material parameters – **Table 1**). Then, the FE solver returns the max. nodal computed values representing output parameters such as: the normal stress, displacement, and von Mises stress. These are then plotted in subsequent sections to describe the resultant Poynting effect. Thus, to eliminate redundancies, all constitutive equations are described without any material phase subscripts ($\boldsymbol{W_{axon}}$ or $\boldsymbol{W_{ECM}}$). However, it is implied that numerically the solver performs these computations for both axons and ECMs mesh elements. Hence, for simplicity, analytical model equations are discussed without any material phase subscripts.

## 2.4 Finite Element Model

In the current study, as a proof-of-concept for realizing the Poynting effect in brain white matter, a micromechanical scale Representative volume element (RVE) geometry of unit dimensions (1µm x 1µm x 1µm), packing two axons in ECM material is designed to set up a finite element method (FEM) model to depict Poynting effect in various simple shear configurations. In the proposed FE model, axons are tied to the surrounding ECM.

The proposed model represents axon fibers as straight cylindrical fibers with an average diameter value of 0.44 µm. Since the proposed RVEs have two effective axons packed with a volume fraction of VF = 0.3 to analyze for Poynting behavior. Refer to **Figure 4** for FEM geometrical setup details. The FEM used a mesh of 81520 elements. Hybrid hexahedral elements (C3D8RH: An 8-node linear brick, hybrid, constant pressure, reduced integration, hourglass control). All elements required a linear hybrid formulation due to the hyperelastic material assignment to reproduce the exact incompressibility. To model the axon and ECM material phases, Ogden hyperelastic material model was deployed as discussed in §2.2 and §2.3 to account for hyper-elasticity.

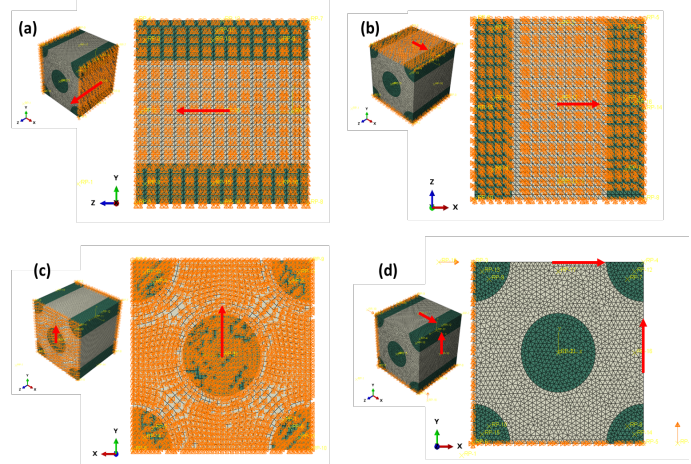

***FIGURE 4:*** *(a) FEM setup describing simple shear configuration along axon fiber (longitudinal shear) boundary conditions for a single RVE FE model, with the back face fixed and a shear stretch applied on the front face. (b) FEM for the simple shear across axonal fibers. (c) FEM setup for simple shear perpendicular to fiber (along the y-axis). (d) Pure shear FEM B.C. setup with shear applied across the axon axis on two RVE faces.*

**FE Model Boundary Conditions**: To simulate a simple shear scenario, one face is fixed (encastered) to depict a face of a tissue sample fixed on a platform. The surface to be sheared in a given direction is pinned in the other two directions to maintain a state of simple shear. A strain value is specified for the corresponding shear model in the ensemble of cases discussed here. A reference point (RP) was defined at the center of the face. The surface subjected to simple shear has the RP that is kinematically coupled to all the points on the edge nodes of the RVE face. The linear shear strain value is defined at last and solved up to 100% shear strain scenario. The field output values of particular interest are stress (S) and displacement (U) field variables.

As mentioned, simple shear cases are solved for three configurations and a fourth additional case of pure shear is also studied to understand the degree of the Poynting effect. For depicting Poynting behavior, the stress field output variable (S11, S22 or S33) and displacement field variable (U1, U2 or U3) depend on the RVE setup acting along an axis perpendicular to the direction of shear is of major significance. These values will be plotted and compared to declare whether RVE FEM shows Poynting behavior for the simple shear ensemble cases.

## 3. RESULTS AND DISCUSSION

In this paper, the developed micromechanical FEM is subjected to three simple shear scenarios which are different in terms of their orientation to axon (fiber) axis. These cases are : 1. Simple shear along axon (shear along z-direction), 2. Simple shear across the axon fiber tracks (i.e., shear along X-direction) and 3. Simple shear perpendicular to the axon fibers (i.e., shear along Y-direction). A fourth case of pure shear is also experimented with to investigate the extent of Poynting behavior. As a representative case, FEM contour plot for single-RVE shear along axon fiber and ECM material combination is shown for Static, General FE simulation step case. The shear moduli for axon and ECM derived from **Table 1** as discussed in §2. **Figure 6** shows the pure shear FEM contour plots for stress distribution for 20% shear strain applied. The difference in stiffness between axon and ECM phases and the inherent nonlinear hyperelastic material characteristics lead to manifestation of Poynting behavior.

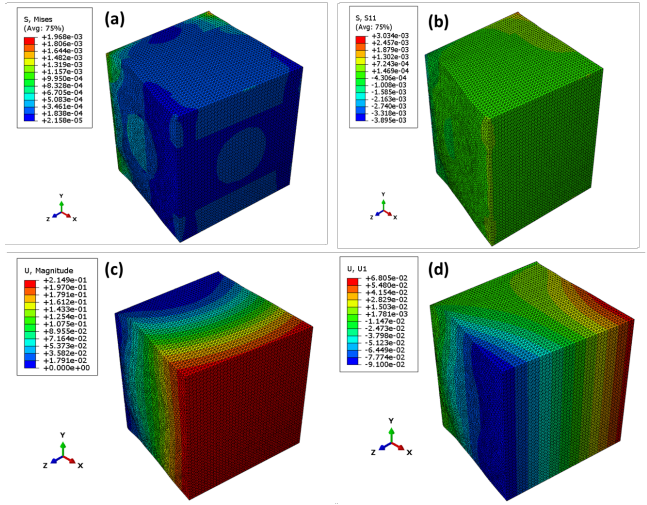

***FIGURE 5:*** *(a) Von Mises stress contour (S) for the axons and the ECM at 20 percent applied stretch for single RVE FEM (shear along axon fiber axis – longitudinal shear). (b) Normal stress (S11) component perpendicular to the direction of applied shear (along the z-axis). (c) Total deformation (U) plot at 20% applied shear strain. (d) Directional displacement (U1) normal to applied shear strain plane depicting positive Poynting behavior.*

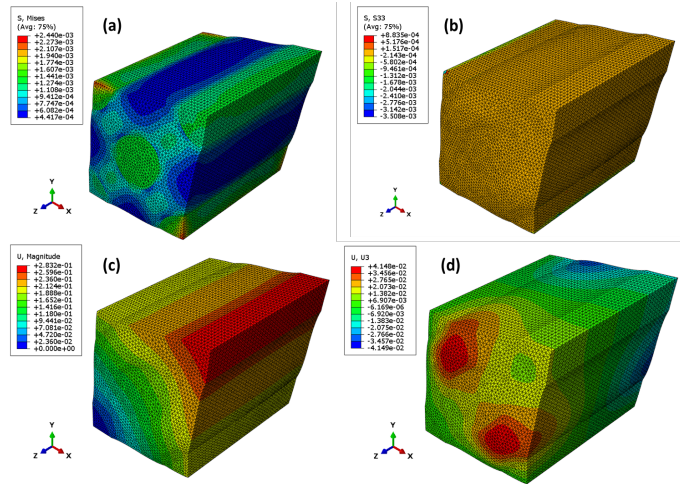

***FIGURE 6:*** *(a) Von Mises stress contour (S) for the axons and the ECM at 20 percent applied stretch for single RVE FEM (pure shear configuration). (b) Normal stress (S33) component perpendicular to the direction of applied shear on the pair of opposite faces across axon fibers. (c) Total deformation (U) plot at 20% applied pure shear strain. (d) Directional displacement (U3) normal to applied pure shear strain planes depicting the modest positive Poynting effect.*

Comparing normal stress and deformation profiles revealed that Poynting behavior was predominant for simple shear cases along axon fibers at lower strain limits. At 20% strain, the pure shear RVE model yielded normal shear at 36.2% and 19.6% of Von-mises (S) and total deformation (U), respectively, clearly indicating the presence of some Poynting behavior. The difference in the shear moduli of the axon and ECM phases contributed to the degree of deformation, leading to the Poynting effect in the bi-phasic model (see **Figure 5**).

### 3.1 Simple Shear along fiber – RVE FEM model

For simple shear along axons fiber axis (Case I) RVE configuration, **Figure 7** shows the stress v/s strain profiles. This model is based on the boundary conditions described in §2.4 (see **Figure 4(a)**). Simple shear is applied on the $XZ\ surface$. The stress is normal to the direction of shearing (S11) and is monitored to describe Poynting behavior.

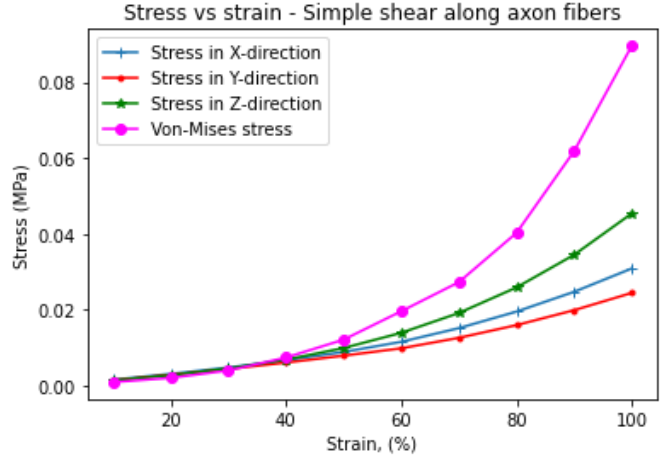

**FIGURE 7:** *Stress (σ) versus strain plot single-RVE FEM for simple longitudinal shear. Impact of axonal direction (FEM geometrical configuration) on resultant Poynting Effect (PE) is analyzed (Case I) and subsequent RMSD analysis done on the results to quantify relative variations in stress profiles v/s von-Mises stress to quantify PE behavior.*

At 100% strain (i.e., 45 deg simple shear strain) [31], maximum value of the positive normal stress S11 is 34.5% of the max. Von-mises stress (S) is observed and max. directional deformation normal to shear direction (U1) is 10.5% of the max. total deformation (U) magnitude, indicating modest (see **Figure 8**) Poynting behavior when sheared longitudinally along axon fiber axis (which is oriented along z-axis).

In each of the analyzed RVE cases, RMSD analysis is conducted to calculate variation in stress profile, von-Mises v/s normal stress component perpendicular to shear planes in all four shear RVE models. Root-Mean-Square Deviation (RMSD) is defined as: $\sqrt{\sum \frac{(f(x_i) - g(x_i))^2}{N}}$ for curves $f(x)$, $g(x)$ and $N$ being number of points $x_i$ at which curves are compared. RMSD b/w von-Mises (S) and normal stress (S11) for Case 1 (*simple shear along axon fibers*) is found to be 0.03022. Meanwhile, RMSD b/w S11 and S22 curves (refer **Figure 7**) is calculated as 0.00384 showing that in both transverse directions to shear planes the normal stress profiles are similar in magnitude.

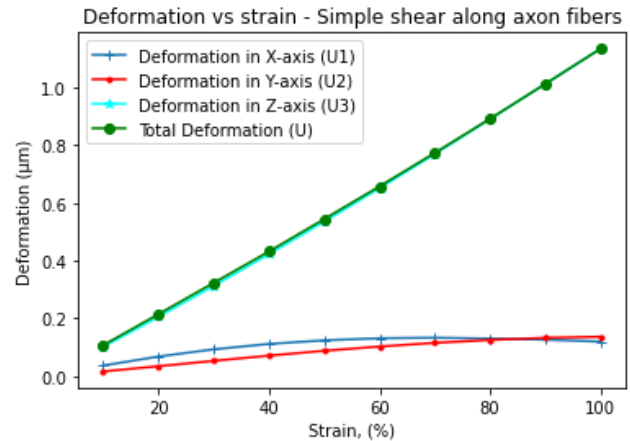

**FIGURE 8:** *Deformation (U, U1, U2, U3) versus shear strain plots for single-RVE FEM, simple longitudinal shear (Case I). The impact of axonal direction (FEM geometrical configuration) on the resultant Poynting Effect (PE) is analyzed (Case I), and subsequent RMSD analysis is done on the deformation results (U1) to quantify relative variations in resultant positive PE behavior with varying strain values.*

RMSD analysis b/w the deformation plots U3 (shear direction) and U1 (direction normal to shear) is 0.7535. Similarly, RMSD b/w U1 and U2 were calculated to be 0.0355, indicating that a similar degree of normal deformation was observed in both transverse directions to shear planes. With increasing simple shear strains, the normal deformation component remained fairly constant (as shown in **Figure 8**), denoting mild Poynting behavior in longitudinal shear configuration.

### 3.2 Simple Shear across fiber – RVE HE model

As in case II, simple shear is applied to the single-RVE FEM across the fiber axis (i.e., the simple shear strain applied along the x-direction on the *XY surface*), **Figure 9** shows the stress v/s strain profiles. This model is based on the boundary conditions described in §2.4 (see **Figure 4 (b)**). The S22 stress component is normal to the direction of shearing and indicates the extent of Poynting behavior. At 100% strain, max. S22 is 59.6% of the max. Von-mises stress (S) and max. Positive directional deformation (U2), which is normal to the direction of simple shear (along the x-axis), is 20.6% verifying the presence of Poynting behavior in the bi-phasic brain white matter (refer **Figure 10**). The second simple shear configuration clearly described greater Poynting behavior than Case 1 (§3.1). This may be attributed to the fact that fibers in this simple shear model are also subjected to some degree of stretching in the transverse direction, and hence a greater magnitude of normal stress would be required to maintain the state of simple shear experimentally. This is in line with the findings of *Destrade et al*. [32], who posited that soft tissues could have a dominant directional Poynting behavior.

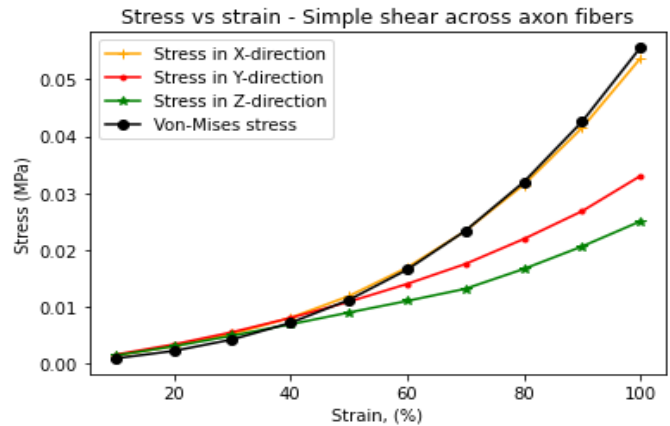

**FIGURE 9:** *Stress ($\sigma$) versus strain plot single-RVE FEM for simple transverse shear (across axon fibers). The impact of axonal direction (FEM geometrical configuration) on the resultant Poynting Effect (PE) is analyzed (Case II). RMSD analysis showed more even distribution between normal stress profiles v/s von-Mises stress and greater PE behavior compared to Case I.*

RMSD b/w von Mises (S) and normal stress (S22) for Case II (*simple shear across axon fibers – transverse shear*) is found to be 0.01227. This is less than the RMSD obtained in Case I, demonstrating comparatively greater Poynting behavior. Meanwhile, RMSD b/w S22 and S33 curves (refer to **Figure 9**) are calculated as 0.00522, implying that normal stresses in the transverse direction to shear planes are similar in magnitude.

RMSD b/w deformation plots U1 (shear direction) and U2 (direction normal to shear) are 0.6264 (**Figure 10**). The decreasing RMSD value clearly describes the greater relative positive Poynting effect (*compared to Case I*). RMSD b/w U3 and U2 calculated to be 0.1543, verified that transverse shear led to a dominant stretching across axonal fibers (along y-axis) and hence more obvious positive directional deformation (Poynting behavior).

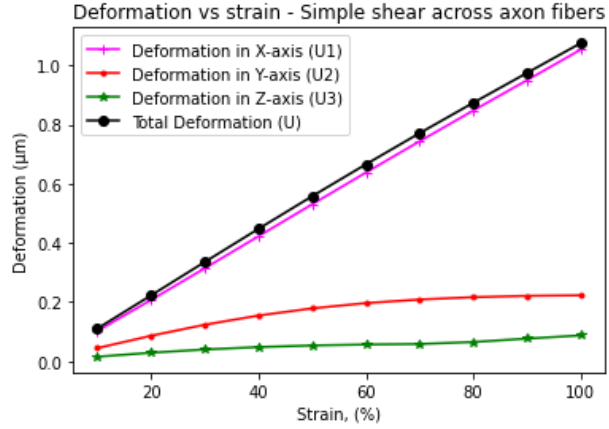

**FIGURE 10:** *Deformation (U, U1, U2, U3) versus shear strain plots for single-RVE FEM, transverse simple shear (Case II). RMSD analysis done on the deformation results (U2 vs U to quantify relative variations in resultant positive PE behavior with increasing shear strain.*

### 3.3 Simple Shear Perpendicular to fiber –HE model

For the final simple shear model (Case III), simple shear is applied to the single-RVE FEM perpendicular to the fiber axis (i.e., the simple shear strain applied along the y-direction on the $YZ\ surface$), **Figure 11** plots the stress v/s strain trends for this setup. In this FEM, as per the boundary conditions defined in §2.4 (see **Figure 4 (c)**). The S33 stress component is normal to the direction of shearing and indicates the extent of Poynting behavior. At 100% strain, max. S33 is calculated to be 43.95% of the max and observed von Mises Stress (S). A more evenly distributed stress profile is observed along all axes in this configuration. S22 curve (red curve in **Figure 11**) also proved the existence of directional stretch in axonal fibers when sheared perpendicular to the fiber axes. This corroborates with the theoretical models put forward by Destrade et al. [32], which underscores the directional dependence of the Poynting effect in sheared brain white tissues.

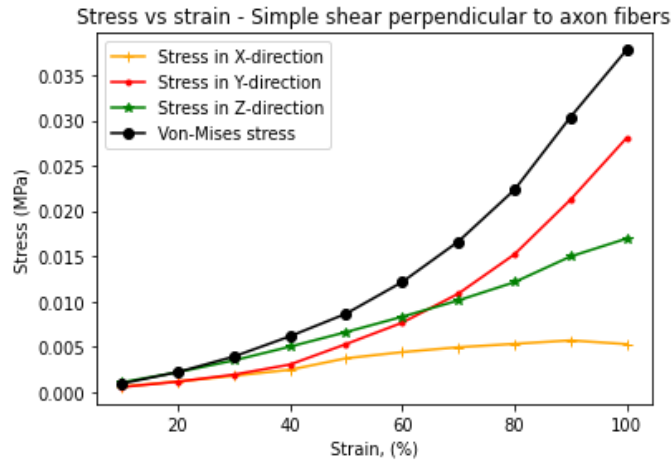

**FIGURE 11:** *Stress (σ) versus strain plot single-RVE FEM for simple perpendicular shear (normal to axon fibers axes). The impact of axonal direction (FEM geometrical configuration) on the resultant Poynting Effect (PE) is analyzed (Case III). Axonal fiber stretching contributed to significantly higher normal stresses. Perpendicular stresses (S33) normal to the shear plane (green curve) also suggest greater PE behavior than Case I.*

RMSD b/w von Mises (S) and normal stress (S33) for Case III (*simple shear perpendicular to axon fibers*) is found to be 0.01206. It is less than the RMSDs obtained in Case I & II (stress profiles), demonstrating the greatest

Poynting behavior amongst three simple shear models. It was also validated by comparing max. normal stress values were 8.64% higher in Case III compared to Case II. RMSD b/w S11 and S33 curves (refer to **Figure 11**) is calculated as 0.00673, verifying that the y-axis is the prominent transverse direction exhibiting the Poynting effect [15, 27].

RMSD b/w deformation plots U2 (shear direction), and U3 (direction normal to shear) is 0.7556. There is a slightly higher RMSD value between deformation curves than in Case II. The positive Poynting effect is still exhibited in Cases I & *II*. Deformation analysis (**Figure 12**) revealed that the normal deformation component (U3) is 9.55% of the max. Deformation (along the y-axis) in Case III. Thus, describing a modest Poynting effect. Although, RMSD b/w U1 and U3 were determined to be 0.05323, suggesting that contribution from higher order invariants by incorporating other material models (such as transverse isotropic or anisotropic hyperelastic) can be explored to determine more perceptible positive Poynting behavior in Case III configuration [32]. Such models would require traction forces to be applied on the side faces to maintain simple shear; hence, more pronounced directional positive or negative deformation depicting Poynting behavior could come to light [9]. Nevertheless, the proposed P.O.C. FEM stress profiles confirm Poynting behavior in Case III.

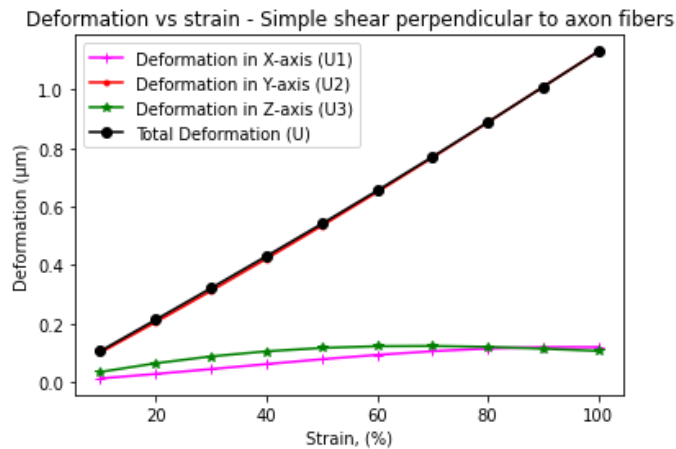

***FIGURE 12:*** *Deformation (U, U1, U2, U3) versus shear strain plots for single-RVE FEM, perpendicular simple shear (Case III). RMSD analysis done on the deformation results (U3) vs U to quantify variations in resultant PE behavior with increasing shear strain.*

### 3.4   Pure shear model – HE model

While many research studies have investigated Poynting effect for twisting of cylindrical white matter tissues, few numerical models have explored numerical realization of pure-sheared rectangular brain matter specimens for Poynting effect. In this regard, a fourth case was analyzed to model single-RVE subjected to pure shear boundary conditions (B.C.). In this FEM setup, shear strains are applied along x- and y-directions on the $YZ\ and\ XZ\ surfaces$) respectively. **Figure 4(d)** depicts the FEM B.C. for pure shear RVE analysis. **Figure 13** plots the stress v/s strain trends. For the pure shear model, at strains close to 60% the stress profiles increase dramatically, and this could be attributed to stress concentrations on the stretched axonal fibers when subjected to pure shears. This exponential rise can also be due to numerical factors whereby the maximum von-mises for axonal node elements act as stress concentration zones due to some degree of bending that occurs in the fibers.

When analyzed for strains beyond 60%, the single-RVE pure shear FEM model suffered convergence issues due to excessive distortions in the hybridized FEM elements. However, as past analytical models from literature and experimental studies hypothesized, solving for strain up to 60% did imply that the proposed model with linearized hybrid elements could depict previously attained extension limits computationally [10] on brain tissues to describe the Poynting effect. In the future, the present single-term Ogden model can be expanded to include the high-order invariants, and computationally high-order (quadratic hybrid FEM elements) may be explored to check model efficacies for extreme pure shear strain limits.

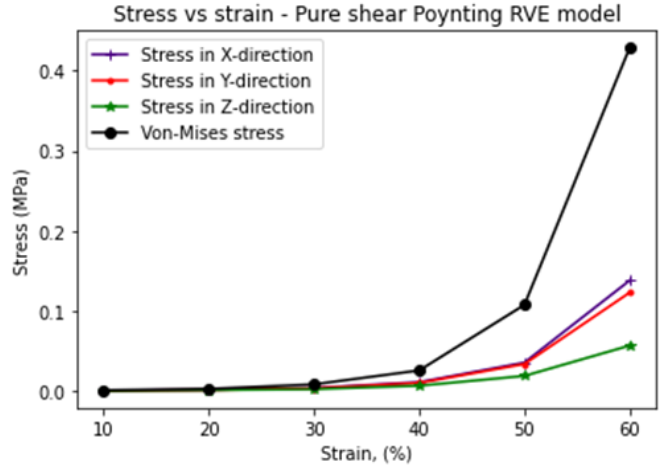

**FIGURE 13:** *Stress (σ) versus strain plot single-RVE FEM for pure shear (strain on pair of faces across fibers - Case IV). Some degree of axonal bending contributed to an exponential rise in normal stresses. Perpendicular stresses (S33) normal to shear planes proved the existence of positive Poynting behavior in pure-sheared brain white matter.*

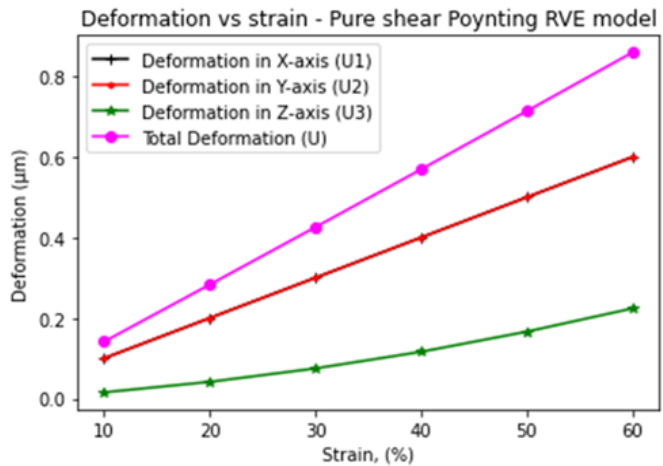

**FIGURE 14:** *Deformation (U, U1, U2, U3) versus shear strain plots for pure shear strained single-RVE FEM (Case IV). RMSD analysis was done on the deformation results (U3) vs. U to quantify variations in resultant PE behavior with increasing shear strain.*

From **Figure 13**, the max. The stress analysis revealed that the normal stress component (S33) to planes of pure shear was 13% of the max. recorded von Mises stress values. The high RMSD between the von Mises and normal stress curves is due to the high shear stress terms contributing to the overall von Mises magnitude. Tensile directional deformation (U3) revealed 26% of the total deformation (U) in this case (**Figure 14**) and confirmed the presence of positive Poynting behavior in pure shear RVEs.

RMSD b/w von Mises (S) and normal stress (S33) for Case IV (*pure shear model*) is found to be 0.15631 (Refer **Figure 13**). It is the highest RMSDs obtained for stress profiles, which could be mainly due to computational reasons on excessively distorted hybrid pure-sheared elements. RMSD b/w S11 and S33 were observed to be 0.03394, underscoring that when subjected to pure shear fiber also experiences stretching, and there is still a consequent normal stress (S33) normal to pure shear planes describing net positive Poynting behavior.

RMSD b/w total deformation (U) and U3 curves (refer **Figure 14**) are calculated as 0.42967, the least RMSD among all four analyzed configurations, indicating the greatest relative Poynting behavior at higher strains up to 60% due to stress concentrations arising in bending axons subjected to pure shear. It is also supported by lower RMSD value (0.26345) b/w pure shear and normal deformation (U3) verifying enhanced resistance to shear in pure shear condition.

### 3.5 Comparison of Poynting behavior (all cases)

Manifestation of Poynting effect in nonlinear materials is dependent on geometrical factors, material parameters and it is also highly influenced by nature of the shear or twist applied [6]. Hence, a comparative numerical analysis is merited to understanding how Poynting effect varies with different shear load cases when analyzed using a single-RVE FEM setup. In this section, all the four previously discussed cases are compared to quantitatively describe variation in Poynting effect in bi-phasic brain white matter FE model. For this, the RMSD analyses is performed to determine variations in Poynting behavior in terms of normal stress and normal component of deformation.

The Poynting behavior in nonlinear models are often characterized or modeled by two major parameters. One of the metrics being the presence of stresses normal to direction of applied shear and another is positive or negative deformation normal to direction of shear. RMSD analysis for the stress profiles for all four analyzed cases yielded that in terms of stress Poynting behavior increased as follows: Case II < Case III < Case I < Case IV (pure shear) FEM model. RMSD between Case I and Case III, 0.00184 < Case I and Case III, 0.00297< Case I and Case IV, 0.01887 (refer **Figure 15**). Hence, all simple shear models show similar degree of Poynting behavior. But, excessive bending and shear distortions buildup in pure shear model meant that Case IV requires a much higher amplitude of normal stress component to maintain a state of pure shear in brain white matter, i.e., indicating the greatest degree of Poynting effect among all analyzed models.

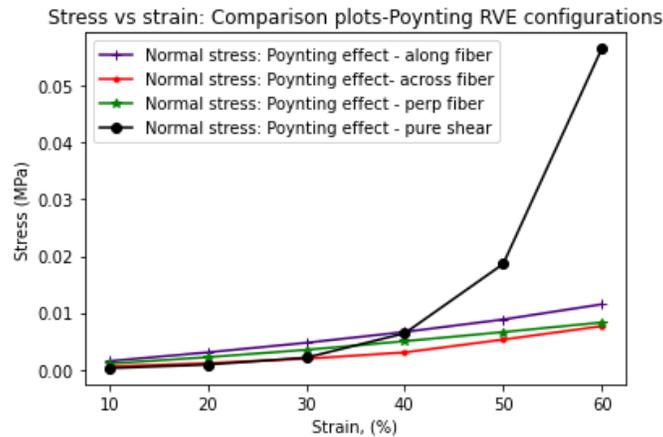

**FIGURE 15:** *Stress ($\sigma$) versus shear strain % plots of single-RVE FEM comparing the extent of demonstrated Poynting effect in all four shear configurations (Case I -IV). Excessive distortions led to high-stress concentration in the pure shear model (black curve). All four models verified the existence of positive Poynting behavior in bi-phasic isotropic HE brain white matter.*

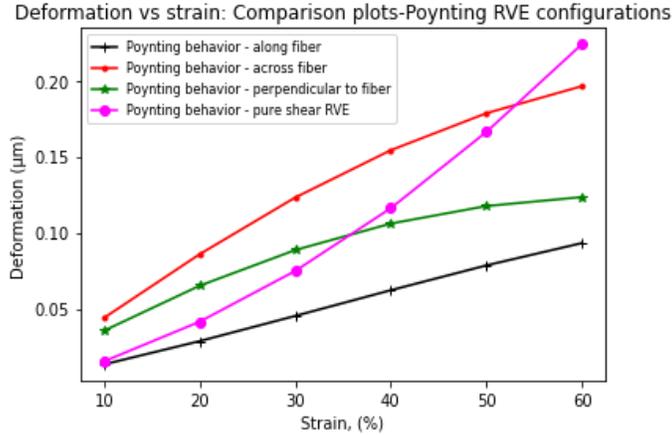

**FIGURE 16:** *Deformation (U, U1, U2, U3) versus shear strain % plots for all four single-RVE FEM models (Case I-IV). RMSD analysis revealed a transitional trend in the extent of positive Poynting behavior, with a pure shear model due to bending stresses in axons at higher strains.*

RMSD analysis for the deformation plots showed a transitional behavior in the degree of the Poynting effect. In **Figure 16**, RMSD between Case I (black curve) and Case III (green curve) is 0.0054 < RMSD b/w Case I and Case II (red), 0.0412 < RMSD b/w Case I and Case IV / pure shear model (pink), 0. 04437. It is clear to see that at lower strain limits, simple shear applied across fiber and perpendicular to axonal fibers showed a dominant Poynting effect, but as the strains were increased to 60%, excessive shear deformations in pure shear RVE model led to a pronounced positive Poynting displacements along the z-axis (U3, as discussed in **§3.4**).

## 4. CONCLUSION

Shearing of nonlinear tissues is not a simple phenomenon; it is often accompanied by normal forces resulting from material non-linearities. This coupling of normal and shear tractions for brain white matter (axon and glial tissues) leads to the Poynting effect manifestation. From a physiological and biomedical standpoint, numerical modeling of the Poynting effect in non-linearly behaving soft tissues has found application in surgical planning and tissue-surgical tool interaction simulations [27]. The results obtained from this proposed computational study will help propel research in nonlinear soft tissue characterization and quantitatively investigate the existence & degree of Poynting behavior when subjected to the ensemble of shear strains.

The current study, a proposed 3D FEM framework is developed for a single-RVE setup to simulate simple shear and pure shear strain scenarios in two-fiber packing bi-phasic (axon and ECM) RVEs. A series of simulations are carried out for three simple shear FE models, and a fourth pure-shear model is analyzed to depict the Poynting effect using a Steady state (general, static) setup in Abaqus. For simple shear models, Ogden HE modeled brain white matter showing evidence of the Poynting effect in all analyzed configurations. A positive Poynting effect (tissue expanding in the direction normal to shear direction) was observed for single-RVE FEM. At lower strains, longitudinally sheared axons (Case I) showed a greater degree of Poynting behavior, but as strain limits were gradually increased, shear deformation buildup in the pure shear model (Case IV) exhibited the highest directional deformation normal to shear planes. The results from the single-RVE model also proved that loading configuration does affect degree of Poynting effect as in Case II and Case III axonal fibers undergo some degree of stretching and hence greater Positive Poynting behavior (refer **Figure 16**). In all analyzed cases, numerical results suggested that specimen thickness was seen to have no influence on shear stresses for an isotropic Ogden HE modeled brain white matter RVE. Although, if more RVEs are joined together, then stresses are expected to scale-up. This is in line with numerical studies conducted by *Rashid et. al* [10] who observed similar thickness independence for homogenized brain matter model subjected to shear loads.

Proposed proof-of-concept micromechanical FEM has potential limitations. First, the model approximates tie-contact between axon and ECM phases, even though physiologically they exhibit more transitional kinematics between them [13, 20, 24]. FE models with surface traction interactions defined between axon and ECM is being developed and could yield models with greater fidelity [15]. In this model, only one single-RVE FEM was presented as P.O.C., in future scalable RVE models will be examined to analyze shear stress buildup. In the future, more 3D

RVE models will be developed with inbuilt axonal tortuosity to determine its effect on Poynting behavior. This model would help evaluate cerebral damage mechanics and its dependence on axonal geometry, varying brain mass, shear loading magnitude, and strain rates by incorporating Hyper-viscoelastic material properties in the current POC FEM's material model. Parameterization of fiber packing will be performed in future test cases to understand stress buildup in densely packed RVEs subjected to simple and pure shear strains. Also, the authors propose that aging in brain matter can be computationally characterized by tuning the volume fractions (V.F.) in the RVEs for myelinated axons to analyze the effect of shear loads on hyper-elastically defined decaying or injured brain matter [12, 33]. While many previous studies have proposed mathematical or macroscopic (homogenized) computational modeling of Poynting behavior, this study is a novel attempt to help bridge the gap between macro-scale and microscale shear mechanics for brain matter and pave the way for transitional multi-scale characterization of Poynting effect in scalable brain tissue FEM.

Computationally feasible and elaborate 3D FEM explicit dynamics model with time-domain hyper-viscoelastic models could offer tremendous insights into damage initiation and bending stress in RVEs with tortuous axonal tracks. Two or three-term Ogden HE material models could be explored to numerically simulate contributions of high-order invariants to validate dominant positive or negative Poynting behavior numerically [32] of different soft tissues.

**ACKNOWLEDGEMENTS**

NSF Grants CMMI-1436743, CMMI-1437113, CMMI-1762774, CMMI-1763005 provided support.